\input{aipcheck}

\documentclass[
    ,final            % use final for the camera ready runs
  ]
  {aipproc}

\layoutstyle{6x9}

\begin{document}

\title{Spherically symmetric models: 
 separating expansion from contraction  in models with anisotropic pressures}

\classification{98.80.-k, 98.80.Cq, 98.80.Jk, 95.30.Sf , 04.40.Nr, 04.20.Jb 
%choose from this list          \texttt{http://www.aip..org/pacs/index.html}
}

\author{Jos\'{e} P. Mimoso}{
  address={Physics Department, Faculty of Science \& \\
  Centro de Astronomia e Astrof\'{\i}sica da Universidade de Lisboa
\\
 Faculdade de Ci\^encias, 
 Ed. C8, Campo Grande, 1769-016 Lisboa, Portugal}
}

\author{Morgan Le Delliou}{
  address={Centro de Astronomia e Astrof\'{\i}sica da Universidade de Lisboa
\\
 Faculdade de Ci\^encias, 
 Ed. C8, Campo Grande, 1769-016 Lisboa, Portugal}
}

\author{Filipe C. Mena}{
  address={Centro de Matem\'{a}tica, 
 Universidade do Minho\\
 Campus de Gualtar, 4710-057 Braga, Portugal}
  %,altaddress={<author1 address>} % additional visiting address
}

\begin{abstract}
We investigate spherically symmetric spacetimes with an anisotropic fluid and
discuss the existence and stability of a dividing shell separating 
expanding and collapsing regions.  We find that the dividing shell is defined by a relation between the pressure gradients, both isotropic and anisotropic, and the strength of the fields  induced by the Misner-Sharpe mass inside the separating shell and by the pressure fluxes. This balance is a generalization of the Tolman-Oppenheimer-Volkoff equilibrium condition which  defines a local equilibrium  condition, but conveys also a non-local character given the  definition of the Misner-Sharpe mass. We present a particular  solution with dust and radiation that provides an illustration of our results.

 % This template file shows how to use the \texttt{aipproc} class to
 % produce a paper with the correct layout for \emph{%
 %  AIP Conference Proceedings  6in   x 9in single column}.

 % A full description of the features supported by the \texttt{aipproc}
 % class can be found in the \texttt{aipguide.pdf} document accompanying
 % the distribution.

 % Frequently asked questions can be found in the \texttt{FAQ.txt}
 % document.
\end{abstract}

\maketitle

%%%%%%%%%%%%%%%%%%%%%%%%%%%%%%%%%%%%%%%%%%%%
%% MAINMATTER
%%%%%%%%%%%%%%%%%%%%%%%%%%%%%%%%%%%%%%%%%%%%

%\section{Introduction}
The universe close to us exhibits structures below certain scales that seem to be immune to the overall expansion of the universe.  This reflects two different gravitational behaviors, and usually the dynamics  corresponding to the  structures that have undergone non-linear collapse is treated under the approximation that Newton's gravitational theory is valid without residual acceleration. However, this approach does not tell us with exactitude what is the critical scale where the latter approximation starts to be valid, nor does it explain in a non-perturbative way how the collapse of over-dense patches decouples from the large-scale expansion. For this purpose one requires a fully general relativistic approach where  an exact solution exhibiting the two competing behaviors and allowing us to characterize how the separation between them comes about. It is the understanding of this interplay between collapsing and expanding regions within the theory of general relativity (GR)
   that we aim to address here.

In previous works we have investigated the present issue in models with spherical symmetry and with a perfect fluid  \cite{Mimoso et al 2010,Le Delliou et al 2011} . Here we briefly report our findings when we overcome the limitations of  a perfect fluid description of the non-equilibrium setting under focus. We thus consider here an anisotropic fluid. 

We resort   to a $3+1$ splitting,  and assess the existence and stability of a dividing shell separating expanding and collapsing regions, in a  gauge invariant way.

%\section{Conditions for the existence of a separating shell}
In the generalized Painlev\'e-Gullstrand coordinates~\cite{Lasky:2006zz}  spherically symmetric metrics can be cast as
\begin{equation}
ds^{2}=-\alpha(t,r)^{2}dt^{2}+\frac{1}{1+E(t,r)}\left(\beta(t,r)dt+dr\right)^{2} 
+R^{2}(r,t)\,d\Omega^{2}.\label{eq:dsLaskyLun-1} \; .
\end{equation}
In the latter expression $\alpha(t,r)$ is the lapse function and $\beta(t,r)$ the shift function. Notice that the areal 
radius $R$ differs, in principle, from the $r$ coordinate to account for additional degrees of freedom that are required to cope with both a general fluid that includes anisotropic stresses and heat fluxes. 

We consider an energy-momentum tensor 
\begin{equation}
T^{ab} = \rho \, n^a n^b + P\, h^{ab} + \Pi^{ab} +2 j^{(a} n^{b)}\; , \label{anisot_stressEMT}
\end{equation}
where $n^a=\alpha^{-1}(1,\beta,0,0)$ is the flow vector, $h^{ab}= g^{ab}+n^{a}n^{b}$ is the metric of the hypersurfaces orthogonal to it, $\rho$ is the energy density, $P$ is the pressure, $\Pi^{ab}$ is the anisotropic stress tensor and $q^a$ is the heat flux vector. $\Pi^{ab}n_b=0$ and ${\Pi^a}_a=0$, i.e., the anisotropic stress $\Pi^{ab}$ is  orthogonal to $n^a$ and traceless, and $j^a = j(t,r)(\beta,1,0,0)$ represents the heat flux which is also orthogonal to the matter flow.

Introducing the Misner-Sharpe mass $M$ and following \cite{Lasky:2006zz,Mimoso et al 2012}
\begin{equation}
\frac{\partial}{\partial r}M = 4\pi\left(\rho \frac{\partial}{\partial r}R+j\mathcal{L}_{n}R\right)R^{2} \label{MS_mass}
\end{equation}
it is possible to derive from the Einstein field equations expressions for $\left({\mathcal{L}_{n}R}\right)^2$ and for ${\mathcal{L}_{n}^2 \, R}$. The simultaneous vanishing of the latter quantities defines, on the one hand, the turning point condition ($\left({\mathcal{L}_{n}R}\right)^2=0$) and, on the other hand, the generalized local conditions for the existence of a separating shell:   (${\mathcal{L}_{n}^2 \, R}=0$). This yields
\begin{equation}
\left({\mathcal{L}_{n}R}\right)^2 = \frac{2M}{R}+ (1+E)\,\left(\frac{\partial R}{\partial r}\right) ^2-1 = 0 \label{cond1_TP}
\end{equation}
and
\begin{equation}
-{\mathcal{L}_{n}^2 \, R} = \frac{M}{R^2}+ 4\pi(P-2\Pi)R- \frac{1+E}{\alpha}\,\frac{\partial \alpha}{\partial r}\frac{\partial R}{\partial r}  =0 \; . \label{gTOV2}
\end{equation}
This allows us to extend the generalization of the TOV function made in \cite{Mimoso et al 2010,Le Delliou et al 2011}, which we called gTOV,  to the case where anisotropic  stresses are present, since $
%\begin{equation}
\mathrm{gTOV} = -{\mathcal{L}_{n}^2 \,R} \; 
%\end{equation}
 $. 
 In what follows case we shall ignore the heat fluxes (it is then possible to restrict to $R=r$, but we will keep it $R(r,t)$ for the sake of generality). We have
\begin{equation}
-\frac{1}{\alpha}\,\frac{\partial \alpha}{\partial r} = \frac{1}{(\rho+P-2\Pi)}\,\left[ \frac{\partial}{\partial r}(P-2\Pi) - \frac{6\Pi}{R}\left(\frac{\partial R}{\partial r}\right)\right]
\end{equation}
and Eqn. (\ref{gTOV2}) becomes
%\begin{equation}
%-{\mathcal{L}_{n}^2 \,r} =\frac{M}{r^2}+ 4\pi(P-2\Pi)r+\frac{1+E}{(\rho+P-2\Pi)}\,\left[ \frac{\partial}{\partial r}(P-2\Pi) - \frac{6\Pi}{r}\right]
%\end{equation}
\begin{equation}
\frac{M}{R^2}+ 4\pi(P-2\Pi)R+\frac{1+E}{(\rho+P-2\Pi)}\,\left[ \frac{\partial}{\partial r}(P-2\Pi) - \frac{6\Pi}{R}\left(\frac{\partial R}{\partial r}\right)\right]\frac{\partial R}{\partial r}= 0 \; . \label{anis_gTOV}
\end{equation}
which is the gTOV$=0$ equation of state for the stationarity of the separating shell, and it is immediately apparent that, when going from the isotropic perfect fluid to the case of an anisotropic content, we have to replace $P$ by $P-2\Pi$ in the equations. This means of course that the anisotropic stresses play a fundamental role in defining the pressure gradients that  promote the local conditions for the separability of the sign of ${\mathcal{L}_{n}R}$.

It is possible to relate ${\mathcal{L}_{n}R}$ with the expansion and shear scalars, respectively, $\Theta=n_{\,;a}^{a}$,
and $a$, where $a$ can be defined from the shear tensor $\sigma_{ij}$ as
%\begin{equation}
$\sigma_{ij}= a(t,r)\,  P_{ij} $
%\end{equation}
where we use the fact that, from the spherical symmetry, all the quantities $X_{ij}={h_i}^a  {h_i}^a\,  X_{ab}$ 
share the same spatial eigendirections characterized by the traceless 3-tensor ${P^i}_j = \rm{diag}\left[-2,1,1 \right]$. We have
\begin{equation}
\left(\frac{\theta}{3}+a\right) = \frac{{\mathcal{L}_{n}R}}{R} \label{H_r}
\end{equation}
and we see that the turning point condition (\ref{cond1_TP}) does imply neither the vanishing of  the expansion nor of the shear, but it rather means that these quantities  should satisfy $\theta_\ast = 3a_\ast$ at the separating shell $r=r_\ast$. If either $\theta$ or $a$ were to vanish at this locus we would then have the other quantity vanishing as well.This limit case corresponds to  a static separating shell.

%\section{Two illustrative applications}
Given  Eqn. (\ref{H_r}) it is  interesting to relate the condition (\ref{cond1_TP}) to the Hamiltonian constraint 
%(\ref{eq:Hamiltonian}) 
that generalizes the Friedman equation
\begin{equation}
\left(\frac{\theta}{3}+a\right)^2= \frac{8\pi \rho}{3} -\frac{{}^3R}{6}+2a\,\left(\frac{\theta}{3}+a\right)  \; .
\end{equation}
We conclude that
\begin{equation}
\frac{2M}{R}+ (1+E)\,\left(\frac{\partial R}{\partial r}\right) ^2-1=\frac{8\pi \rho}{3} -\frac{{}^3R}{6}+2a\,\left(\frac{\theta}{3}+a\right)  \; . \label{local_cond1_TP}
\end{equation}
which allows us to emphasize that the stationarity condition for the existence of a separating shell (\ref{cond1_TP}) involves non-local quantities namely $M$, and $E$, while the right-hand side of the expression just derived, (\ref{local_cond1_TP}), only involves local quantities. The non-locality of the conditions  (\ref{cond1_TP}) and (\ref{anis_gTOV}) is consistent with the findings of Herrera and co-workers \cite{Herrera} who have studied the ``cracking'' of compact objects in astrophysics using small anisotropic perturbations around spherically symmetric homogeneous fluids in equilibrium. 

%In parallel we also wish to clarify the relation between the gTOV function 
%\begin{equation}
%gTOV = \frac{M}{R^2}+ 4\pi(P-2\Pi)R+\frac{1+E}{(\rho+P-2\Pi)}\,\left[ \frac{\partial}{\partial r}(P-2\Pi) - \frac{6\Pi}{R}\left(\frac{\partial R}{\partial r}\right)\right]
%\end{equation} 
%and the "generalized" Raychaudhuri equation
%\begin{equation}
%{\mathcal{L}_{n}\left( \frac{\theta}{3}+a\right)} + \left( \frac{\theta}{3}+a\right)^2 = -\frac{4\pi G}{6}\,(\rho+3p) - \left(\Sigma-\frac{8\pi G}{2}\Pi\right) \; .
%\end{equation}

%Although the conditions (\ref{cond1_TP}) and (\ref{anis_gTOV}) that characterize the separating shell hold  locally, at $r=r_\ast$, they involve  non-local quantities, namely $M$ and $E$. Indeed, from the definition of $M$, Eq. (\ref{MS_mass}) we see that the profile of the distribution of matter inside the separating shell  is taken into account.

We now turn our attention to an exact solution derived by Sussman and Pav\'on for a spherically symmetric model with a combination of dust and radiation exhibiting anisotropic stress, but no heat fluxes \cite{Sussman:1999vx}. The metric is  written in the Lema\^{\i}tre-Tolman-Bondi (LTB) form
\begin{equation}
ds^{2}=-dT^{2}+\frac{\left(\partial_{r}R\right)^{2}}{1+E(T,r)}dr^{2}+R^{2}d\Omega^{2},\label{eq:dsLTB}
\end{equation}
and it assumed that the flow is geodesic to keep ourselves as close as possible to the case where dust is the only component which is present, i.e., the original LTB case. The latter hypothesis implies in GPG coordinates that 
%\begin{equation}
$-\frac{1}{\alpha}\,\frac{\partial \alpha}{\partial r} = 0$, 
%\end{equation}
and hence
%\begin{equation}
$(P-2\Pi)^\prime + 6\Pi \,\frac{R'}{R} = 0$
%\end{equation}
where the prime stands for differentiation with respect to $r$. On the other hand the absence of heat fluxes makes $E=E(r)$, independent of $T$. The condition (\ref{cond1_TP}) amounts to
\begin{equation}
\frac{2M(r)}{R}+\frac{2W(r)}{R^2} +E= 0 \; ,
\end{equation}
and equating the gTOV condition reduces to
\begin{equation}
\frac{M(r)}{R^2}+\frac{W(r)}{R^3} +4\pi(P-2\Pi)R=0  \;.
\end{equation}
The latter conditions defining  the separating shell reduce to a differential equation that $E(r)$ must satisfy, and in \cite{Mimoso et al 2012} we show that it is possible to specify such an $E$.

Thus, in the present work we find that the existence of shells separating expanding and contracting domains of the areal radius is defined by two relations: an energy balance yielding a stationarity condition, and  a relation between the pressure gradients, both isotropic and anisotropic, and the strength of the fields induced by the Misner-Sharpe mass inside the separating shell and by the pressure fluxes. This balance is a generalization of the Tolman-Oppenheimer-Volkoff equilibrium condition which  defines a local equilibrium  condition, but simultaneously has a non-local character given the  definition of the Misner-Sharpe mass M (and incidentally that of the function $E$ as well). We wish also to emphasize that the consideration of anisotropic stresses is most important to guarantee the fulfillment of this gTOV and has an impact of the propagation of the shear. A more detailed and complete discussion of this subject can be found in \cite{Mimoso et al 2012}.

%\subsection{<A subsection>}

%Some url test \url{http://www.world.universe}.

%\subsubsection{<A subsubsection>}

%\paragraph{<A subsubsubsection>}

%%%%%%%%%%%%%%%%%%%%%%%%%%%%%%%%%%%%%%%%%%%%%%%%
%% BACKMATTER
%%%%%%%%%%%%%%%%%%%%%%%%%%%%%%%%%%%%%%%%%%%%%%%%

\begin{theacknowledgments}
The authors wish to thank Jos\'e Fernando Pascual-Sanchez for helpful discussions. FCM is supported by CMAT,
Univ. Minho, and FCT projects PTDC/MAT/108921/2008 and CERN/FP/116377/2010. JPM also wishes to thank FCT for the grants  PTDC/FIS/102742/2008  and  CERN/FP/116398/2010. 
\end{theacknowledgments}

%%%%%%%%%%%%%%%%%%%%%%%%%%%%%%%%%%%%%%%%%%%%%%%%
%%%%%%%%%%%%%%%%%%%%%%%%%%%%%%%%%%%%%%%%%%%%

\bibliographystyle{aipproc}   % if natbib is available

\begin{thebibliography}{9}
\bibitem{Mimoso et al 2010} 
%\cite{Mimoso:2009wj}
%\bibitem{Mimoso:2009wj} 
  J.~P.~Mimoso, M.~Le Delliou and F.~C.~Mena,
  %``Separating expansion from contraction in spherically symmetric models with a perfect-fluid: Generalization of the Tolman-Oppenheimer-Volkoff condition and application to models with a cosmological constant,''
  Phys.\ Rev.\ D {\bf 81}, 123514 (2010)
  [arXiv:0910.5755 [gr-qc]].
  %%CITATION = ARXIV:0910.5755;%%

\bibitem{Le Delliou et al 2011} 
%\cite{LeDelliou:2011wk}
%\bibitem{LeDelliou:2011wk} 
  M.~Le Delliou, F.~C.~Mena and J.~P.~Mimoso,
  %``The role of shell crossing on the existence and stability of trapped matter shells in spherical inhomogeneous \Lambda-CDM models,''
  Phys.\ Rev.\ D {\bf 83}, 103528 (2011)
  [arXiv:1103.0976 [gr-qc]].
  %%CITATION = ARXIV:1103.0976;%%

\bibitem{LaskyLun06b}Lasky, P.D., \& Lun, A.W.C., PRD, 74 (2006)
084013

%\cite{Lasky:2006zz}
\bibitem{Lasky:2006zz}
  P.~D.~Lasky, A.~W.~C.~Lun,
  %``Spherically Symmetric Gravitational Collapse of General Fluids,''
  Phys.\ Rev.\  {\bf D75}, 024031 (2007).
  [gr-qc/0612007].

\bibitem{Mimoso et al 2012} 
  J.~P.~Mimoso, M.~Le Delliou and F.~C.~Mena,
 ``Separating expansion from contraction in spherically symmetric models with an anisotropic fluid,''
In preparation.
%  [arXiv:0910.5755 [gr-qc]].
  %%CITATION = ARXIV:0910.5755;%%

%\cite{Mena:2004ck}
\bibitem{Mena:2004ck}
  F.~C.~Mena, B.~C.~Nolan, R.~Tavakol,
  %``The Role of anisotropy and inhomogeneity in Lemaitre-Tolman-Bondi collapse,''
  Phys.\ Rev.\  {\bf D70}, 084030 (2004).
  [gr-qc/0405041].
  
\bibitem{Herrera} L. Herrera, Phys. Lett. A 165, 206 (1992); A. Di Prisco, L. Herrera, E. Fuenmayor and V. Varela,
{\em Phys. Lett. A} \textbf{195}, 23 (1994); A. Abreu, H. Hernandez,
and L. A. Nunez, {\em Classical Quantum Gravity} \textbf{24}, 4631
(2007); A. Di Prisco, L. Herrera, and V. Varela, {\em Gen. Relativ.
Gravit.} \textbf{29}, 1239 (1997); L. Herrera and N. O. Santos, {\em
Phys. Rep.} \textbf{286}, 53 (1997). 


%\cite{Sussman:1999vx}
\bibitem{Sussman:1999vx}
  R.~A.~Sussman, D.~Pavon,
  %``Exact inhomogeneous cosmologies whose source is a radiation matter mixture with consistent thermodynamics,''
  Phys.\ Rev.\  {\bf D60}, 104023 (1999).
  [gr-qc/9907010].


\end{thebibliography}

%%%%%%%%%%%%%%%%%%%%%%%%%%%%%%%%%%%%%%%%%%%

\end{document}